\documentclass[aps,prl,10pt,twocolumn,superscriptaddress,balancelastpage,showpacs,reprint,english]{revtex4-1}
\usepackage[T1]{fontenc}
\usepackage[latin9]{inputenc}
\usepackage{amsmath}
\usepackage{amssymb}
\usepackage{commath}
\usepackage{graphicx}
\usepackage{color}
\usepackage{bm}
\usepackage[caption=false]{subfig}
\usepackage{mathrsfs}
\usepackage{hyperref}
\usepackage{babel}

\DeclareMathAlphabet{\mathdutchcal}{U}{dutchcal}{m}{n}

\usepackage{tikz}
\usetikzlibrary{decorations.pathmorphing, patterns,decorations.text}
\tikzset{snake it/.style={decorate, decoration=snake}}
\usetikzlibrary{shapes.misc}
\tikzset{cross/.style={cross out, draw=black, minimum
size=2*(#1-\pgflinewidth), inner sep=0pt, outer sep=0pt},
cross/.default={7pt}}
\tikzset{cross2/.style={cross out, draw=black, minimum
size=2*(#1-\pgflinewidth), inner sep=0pt, outer sep=0pt},
cross2/.default={4pt}}

\makeatletter

\newcommand{\qq}{\begin{eqnarray}}
\newcommand{\qqq}{\end{eqnarray}}

\newcommand{\p}{\partial}

\newcommand{\bx}{\mathbf{x}}

\newcommand{\bfx}{\mathbf{x}}
\newcommand{\bfr}{\mathbf{r}}

\newcommand{\bq}{{\mathbf{q}}}
\newcommand{\bG}{\mathbf{\nabla}_{\mathbf{x}}}

\pdfpageheight\paperheight
\pdfpagewidth\paperwidth

\makeatother

\begin{document}

\title{Interface roughening in nonequilibrium phase-separated systems}
\author{M. Besse}
\affiliation{Service de Physique de l'Etat Condens\'e, CEA, CNRS Universit\'e Paris-Saclay, CEA-Saclay, 91191 Gif-sur-Yvette, France}
\affiliation{Sorbonne Universit\'e, CNRS, Laboratoire de Physique Th\'eorique de la Mati\`ere Condens\'ee, 75005 Paris, France}

\author{G. Fausti}
\affiliation{Service de Physique de l'Etat Condens\'e, CEA, CNRS Universit\'e Paris-Saclay, CEA-Saclay, 91191 Gif-sur-Yvette, France}
\affiliation{Max Planck Institute for Dynamics and Self-Organization, 37077 Göttingen, Germany}

\author{M.E. Cates}
\affiliation{DAMTP, Centre for Mathematical Sciences, University of Cambridge, Wilberforce Road, Cambridge CB3 0WA, United Kingdom}

\author{B. Delamotte}
\affiliation{Sorbonne Universit\'e, CNRS, Laboratoire de Physique Th\'eorique de la Mati\`ere Condens\'ee, 75005 Paris, France}

\author{C. Nardini}
\affiliation{Service de Physique de l'Etat Condens\'e, CEA, CNRS Universit\'e Paris-Saclay, CEA-Saclay, 91191 Gif-sur-Yvette, France}
\affiliation{Sorbonne Universit\'e, CNRS, Laboratoire de Physique Th\'eorique de la Mati\`ere Condens\'ee, 75005 Paris, France}

\date{\today}

% ==============================================================================

\begin{abstract}

Interfaces of phase-separated systems roughen in time due to capillary waves. Because of fluxes in the bulk, their dynamics is nonlocal in real space and is not described by the Edwards-Wilkinson or Kardar-Parisi-Zhang (KPZ) equations, nor their conserved counterparts. We show that in the absence of detailed balance, the phase-separated interface is described by a new universality class which we term $\abs{\bq}$KPZ. We compute the associated scaling exponents via one-loop renormalization group, and corroborate the results by numerical integration of the $\abs{\bq}$KPZ equation. Deriving the effective interface dynamics from a minimal field theory of active phase separation, we finally argue that the $\abs{\bq}$KPZ universality class generically describes liquid-vapor interfaces in two- and three-dimensional active systems.% dynamics in active systems undergoing bulk phase separation.

\end{abstract}

\pacs{???}

\maketitle
% ==============================================================================
% ==============================================================================
% ==============================================================================
% ==============================================================================
The roughening of interfaces is among the best-studied problems in statistical mechanics~\cite{krug1997origins,barabasi1995fractal}. Early theoretical investigations~\cite{peters1979radius,plischke1984active,jullien1985scaling} were concerned with the Eden model~\cite{eden1958}, proposed to describe the shape of cell colonies, and with the ballistic deposition model~\cite{family1985scaling}. Kardar, Parisi and Zhang (KPZ)~\cite{KPZ} discovered an important universality class for growing rough interfaces, by adding the lowest order nonlinearity to the continuum Edwards-Wilkinson (EW) model in which height fluctuations are driven by nonconserved noise and relax diffusively~\cite{edwards1982surface}. Despite its fame, KPZ does not describe all roughening interfaces: in the presence of conservation laws, the KPZ nonlinearity can be forbidden and novel universality classes arise~\cite{krug1997origins,sun1989dynamics,caballero2018strong}. 

A notable category of interfaces, that stands out from all these well-known universality classes, are those arising in phase-separated systems, which roughen due to thermally activated capillary waves~\cite{v1908molekular,rowlinson1982molecular,evans1979nature,wertheim1976correlations,weeks1977structure}. There, because of fluxes in the bulk phases, liquid-vapor or liquid-liquid interfaces have nonlocal dynamics in real space~\cite{bray2001interface,shinozaki1993dispersion}. In the absence of fluid motion, small amplitude capillary waves with wavenumber $\bq$ relax by diffusion at a rate $\tau^{-1} = \sigma \abs{\bq}^3$, with $\sigma$ proportional to the interfacial tension~\cite{bray2001interface}. We consider in the following phase separation in $d + 1$ dimensions and 
we denote by $\hat{h}(\bx,t)$ the height of the resulting interface above a 
$d$-dimensional plane. (This notation is standard for interface problems 
such as KPZ.) The dynamics becomes, in Fourier variables $h(\bq,t)$,
\qq
\label{eq:linearLangevin}
\p_t h(\bq,t) = -\sigma \abs{\bq}^3 h(\bq,t) +\sqrt{2D \abs{\bq}} \eta(\bq,t),
\qqq
where $\eta$ is a Gaussian white noise such that
\qq 
\langle \eta(\bq_1,t_2) \eta(\bq_2,t_2) \rangle = \delta(\bq_1+\bq_2) \delta(t_1-t_2)\,.
\qqq
The linear evolution equation (\ref{eq:linearLangevin}) can be written in real space as 
$\p_t \hat{h} = G\star \nabla^2\hat{h} + \eta_R$
 where $\star$ stands for the spatial convolution, $ \langle \eta_R(\bx_1,t_1) \eta_R(\bx_2,t_2)\rangle \sim 2D G(\bx_1-\bx_2)\delta(t_1-t_2)$ and the kernel $G$ is long ranged: $G(\bx)\sim |\bx |^{-d-1}$ at large distances. 

The interface in passive phase-separated fluids must be described, at least in the stationary state, by an equation that respects detailed balance. 
In this case the interface is a subset of degrees of freedom within a thermally equilibrated state of the full, phase-separated, system and hence is itself in equilibrium. The fluctuation-dissipation theorem then means that any nonlinearity that enters Eq.~(\ref{eq:linearLangevin}) is of the form $\abs{\bq} \,\delta \mathscr{F}_I/\delta h_{\bq}$ for some free energy functional $\mathscr{F}_I[h]$. A simple dimensional analysis argument then shows that there exists no nonlinearity correcting Eq.~(\ref{eq:linearLangevin}) that is relevant in the renormalization group (RG) sense for interfaces of dimension $d=1,2$. Hence the stationary dynamics of diffusive ({\em i.e.}, without momentum conservation) interfaces between phase-separated passive liquids is described by mean-field scaling exponents. Dimensional analysis~\cite{tauber2014critical} of Eq.~(\ref{eq:linearLangevin}) then gives $z=3$ and $\chi=(z-d-1)/2$, in terms of which spatial and temporal correlations scale as $\langle \hat{h}(\bfx, t)\hat{h}(\bfx',t)\rangle\sim |\bfx-\bfx'|^{2\chi}$ and $\langle \hat{h}(\bfx, t) \hat{h}(\bfx,t')\rangle\sim |t-t'|^{2\chi/z}$, while the static structure factor $S=\langle|h({\bq},t)|^2\rangle$ scales as $S\sim |\bq|^{-d-2\chi}$ for small $|\bq|$. 

We shall show in the following that this conclusion changes in phase-separated systems that lack detailed balance. Important examples are found in active matter, where elemental units such as self-propelled colloids, bacteria, or cells extract nonthermal energy from the environment and dissipate it to self-propel~\cite{Marchetti2013RMP}. Active liquid-vapor systems are known to display a phenomenology impossible in equilibrium~\cite{gompper20202020}; for example, phase separation arises even in the absence of any attraction among particles~\cite{Tailleur:08,filyABP,stenhammar2014phase,Cates:15}. Moreover, steady-state currents can be present~\cite{tjhung2018cluster}, and phase-separated states can be sustained even when the interfacial tension is negative~\cite{tjhung2018cluster,fausti2021,bialke2015negative,caballero2022activity}. 
In the simplest cases their large-scale interfacial properties are nonetheless similar to those in passive fluids~\cite{Cates:15,Wittkowski14,nardini2017entropy,partridge2019critical,solon2018generalized,gnan2022critical}. In this situation, it was recently assessed on the basis of numerical simulations of particle models~\cite{lee2017interface,sussman2018soft} and field-theoretical analysis~\cite{fausti2021} that capillary wave theory holds. Indeed, some of us have recently shown that small amplitude, long wavelength capillary waves obey Eq.~(\ref{eq:linearLangevin}), although activity changes the capillary interfacial tension (and can make it negative in some parameter regimes, not considered here)~\cite{fausti2021}.

In this Letter we show that when the underlying dynamics breaks detailed balance, as happens generically in active systems, a new universality class describes the roughening of the liquid-vapor interface. This is because a new nonlinearity, which is RGrelevant for interfacial dimension $d\leq 2$, can appear in Eq.~(\ref{eq:linearLangevin}). We study the ensuing equation, which we term $\abs{\bq}$KPZ, by one-loop RG analysis perturbatively for small $\varepsilon=2-d$. We compute the scaling exponents and confirm our conclusions by numerical simulations of the $\abs{\bq}$KPZ equation in $d=1$. We further argue that 
the $\abs{\bq}$KPZ universality class includes the liquid-vapor interface of active systems undergoing bulk phase separation. We do so by considering Active Model B+ (AMB+)~\cite{tjhung2018cluster}, a minimal continuum description of active systems undergoing phase separation, and showing how $\abs{\bq}$KPZ  emerges as the associated equation for the liquid-vapor interface. We thus predict the emergence of a new universality class for the roughening of liquid-vapor interface in two-dimensional active systems ($d+1=2$), the setup in which they are most commonly studied both theoretically~\cite{Cates:15} and experimentally~\cite{Speck:13,liu2019self,van2019interrupted}. 

To assess which nonlinearities might modify Eq.~(\ref{eq:linearLangevin}), let us first recall the symmetries that have to be respected. First, we should impose invariance under rotations and translations, as well as under a shift in the origin of the reference frame, which translates to $h(\bq,t) \to h(\bq,t)+(2\pi)^{d}\delta(\bq)C$ in Fourier space, for any $C \in \mathbb{R}$. Second, the total amounts of liquid and vapor do not change during roughening, implying that the ``total height'' $\int_{\bx} ~\hat{h}$ is constant. %or, in Fourier, that $\lim_{\bq\to0} \p_t h_{\bq}=0$. 
We further assume that chiral symmetry (invariance under $\bfx\to-\bfx $) is not broken. Under these symmetries, the KPZ nonlinearity is forbidden, as it does not conserve the total height. Moreover, the nonlinearities captured in existing models of conserved surface growth, such as  cKPZ~\cite{sun1989dynamics,krug1997origins} and cKPZ+~\cite{caballero2018strong} are found to be RG irrelevant by dimensional analysis. 

Inspecting Eq.~(\ref{eq:linearLangevin}) it appears that $\abs{\bq}$ plays the role of a mobility in the linear description of capillary waves. This thus suggests to consider nonlinearities in the form 
\qq
\label{eq:Ansatz}
\abs{\bq} \int_{\sum_{i=1}^n \bq_i = \bq } g(\bq |\bq_1,...,\bq_n) h(\bq_1)...h(\bq_n),
\qqq
where the integral is over $\bq_1,...,\bq_n$ with the constraint $\sum_{i=1}^n \bq_i = \bq$ (nonlinearities depending on the frequencies $\omega, \omega_1,...,\omega_n$ could be considered as well without changing any conclusion). Note that the prefactor $\abs{\bq}$ ensures that the total height of the interface is conserved. Assuming that $g$ is analytic in all its arguments, and imposing the symmetries mentioned above, the most relevant nonlinearity that can modify Eq.~(\ref{eq:linearLangevin}) is $g(\bq|\bq_1,\bq_2) = \lambda_1 i \bq_1 \cdot i \bq_2 $. This term is relevant for $d< d_c=2$. We are thus led to investigate the following equation, which we term $\abs{\bq}$KPZ:
\qq
\p_t h = -\sigma \abs{\bq}^3 h + \frac{\lambda_1}{2} \abs{\bq} \mathcal{F}[|\nabla \hat{h}|^2] + \sqrt{2D \abs{\bq}} \eta,
\label{eq:qKPZ}
\qqq
where $\mathcal{F}[\cdot]$ stands for the Fourier transform. It is worth noting that Eq.~(\ref{eq:qKPZ}) differs from the KPZ equation endowed with long-range interactions~\cite{medina1989burgers,chattopadhyay1999nonlocal}. 

We next study the $\abs{\bq}$KPZ equation by RG analysis to one loop, perturbatively in $\varepsilon=2-d>0$. Within the Martin-Siggia-Rose formalism~\cite{tauber2014critical}, the action $S$ associated with Eq.~(\ref{eq:qKPZ}) reads
\qq 
\label{eq:bareqKPZ}
S= \int_{Q} \tilde{h}(-Q) \left( G_0^{-1}(Q) h(Q) + A(Q) - D \abs{\bq} \tilde{h}(Q) \right)
\qqq
where $\tilde{h}$ is a response field, $Q = (\omega,\bq)$, $G_0(Q)^{-1} = -i \omega + \sigma \abs{\bq}^3$ and $A(Q) = (\lambda_1/2) \abs{\bq} \int_{Q_1+Q_2=Q} \bq_1 \cdot \bq_2 h(Q_1)h(Q_2)$.

\begin{figure}
\includegraphics[width=0.4\textwidth]{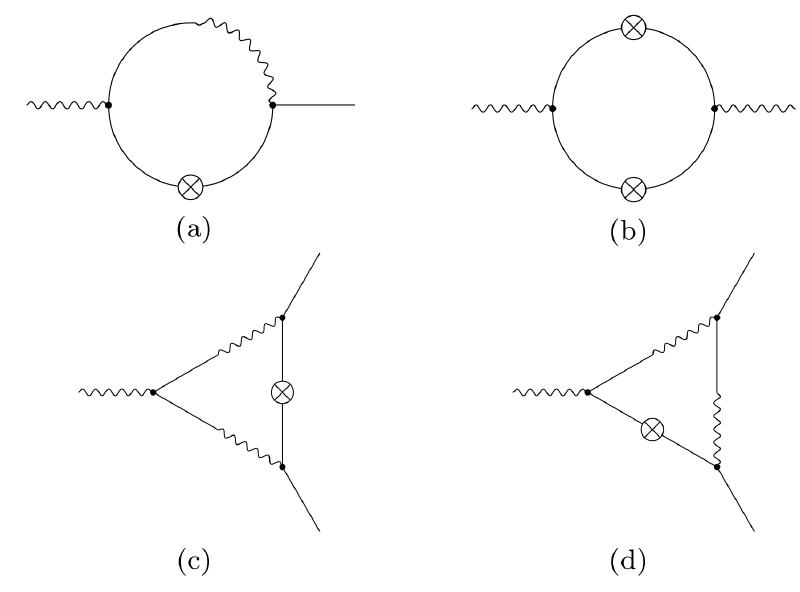}
	\caption{One-loop diagrams for the renormalization of the propagator (a), noise (b) and nonlinearity (c,d).}
	\label{fig:propag}
\end{figure}

We show in Fig.~\ref{fig:propag} the one-particle-irreducible diagrams to one loop associated with the action in Eq.~(\ref{eq:bareqKPZ}) and compute them in~\cite{supp}. The diagram in Fig.~\ref{fig:propag}(a) gives a nonvanishing contribution that renormalizes $\sigma$, while that of Fig.~\ref{fig:propag}(b) only gives irrelevant contributions to the renormalization of the noise. Diagrams in Figs.~\ref{fig:propag}(c)~and~\ref{fig:propag}(d) exactly cancel, as suggested by generalizing the argument of~\cite{janssen1997critical}. (We have confirmed this by explicit computation.) We furthermore show in~\cite{supp} that the $\abs{\bq}$KPZ equation is stable under one-loop perturbative RG flow: any nonlinearity there generated is in the form of Eq.~(\ref{eq:Ansatz}), with $g$ analytic in its arguments, and no linear term more relevant than $\abs{\bq}^3 h(\bq)$ is generated. 

We thereby obtain in~\cite{supp} the following RG flow for the reduced coupling constant $\mathdutchcal{g}=D \lambda_1^2 / \sigma^3$
\qq
\Lambda \frac{\mathrm{d} \mathdutchcal{g}}{\mathrm{d} \Lambda} = \varepsilon \mathdutchcal{g} - \frac{3}{8} K_2 \mathdutchcal{g}^2, \qquad\varepsilon = 2-d.
\label{eq:RG_flow}
\qqq
where $\Lambda$ is the momentum scale, $K_d=S_d/(2 \pi)^d$ and $S_d=2 \pi^{d/2}/\Gamma(d/2)$. 
As expected from dimensional analysis, the Gaussian fixed point ($\mathdutchcal{g}^{\star}=0$)
is the only one of the RG flow for $d\geq 2$: here the mean-field exponents exactly describe the interface properties at large scales. For $d<2$, a new attractive fixed point emerges at $\mathdutchcal{g}^{\star} = 8 \varepsilon /(3 K_2)$. At this fixed point, the scaling exponents are 
\qq\label{eq:critical-exp}
z=3-\frac{\varepsilon}{3},
\qquad \qquad 
\chi = \frac{\varepsilon}{3}\,.
\qqq
These exponents describe the new $\abs{\bq}$KPZ universality class to one loop, that is, to first order in $\varepsilon$. 

To test these predictions, we numerically integrate the $\abs{\bq}$KPZ in $d=1$ using a pseudospectral code with $2/3$ dealiasing procedure. The use of a pseudospectral code is particularly convenient for the $\abs{\bq}$KPZ equation because it has computational complexity $\mathcal{O}(L\log L)$, where $L$ is the system size, while, given its nonlocality in real space, a finite difference code would have complexity $\mathcal{O}(L^2)$. The results presented were obtained with spatial discretization $\Delta x=1$ and time discretization $\Delta t=10^{-2}$.
We quantified the interfacial width via
\qq
W^2(t,L) = \frac{1}{L}\int_{\bx} \hat{h}^2(\bx,t),
\qqq
starting from a flat interface and averaging it over noise realizations. 
Our RG analysis predicts that, while roughening, $W^2\sim t^{2\chi/z}$ and, for a system of finite size $L$, $W^2$ eventually saturates in time to a value $W_\infty^2(L)\equiv W(\infty,L)^2\sim L^{2\chi}$. Measuring the interfacial width as a function of time and its saturated value as a function of $L$ allows us to extract both scaling exponents, $z$ and $\chi$.

\begin{figure}
\includegraphics[width=0.45\textwidth]{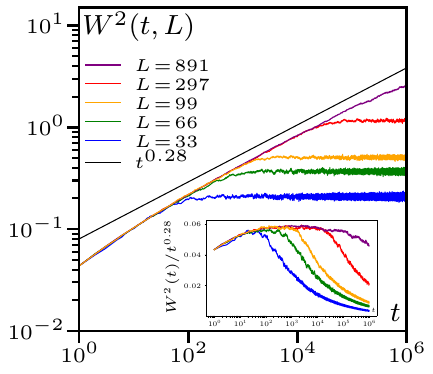}
	\caption{Interfacial width $W^2(t)$ as a function of time for different system sizes showing a roughening law $W^2(t)\sim t^{0.28}$. Parameters used $\sigma=1.0, D=0.1, \lambda=2.0$, corresponding to the reduced coupling constant $\mathdutchcal{g}=0.4$. Each curve was obtained by averaging over $500$ noise realizations. The inset contains the same data but rectified by $t^{0.28}$. 	 }
	\label{fig:W2lambda2}
\end{figure}

\begin{figure}[h!]\center
    \includegraphics[width=0.4\textwidth]{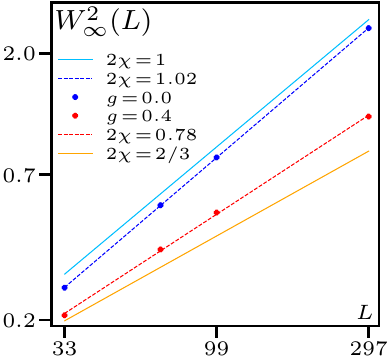}
    \caption{$W_\infty^2(L)\sim L^{2\chi}$ as a function of system size $L$. Continuous lines: RG predictions either at the Gaussian (blue) or $\abs{\bq}$KPZ (red) fixed points. Points: results of numerical integration of the $\abs{\bq}$KPZ equation with $\lambda_1=0$ (blue) and $\lambda_1=2$ (red), corresponding to bare coupling $\mathdutchcal{g}=0.4$. Other parameters: $\sigma=1.0$ and $D=0.1$. Error bars are smaller than the symbols'size. Dashed lines are power-law fits.}
    \label{fig:scalingchi}
\end{figure}

In the case of the linear theory ($\lambda_1=0$) we measure, as expected, $2\chi/z = 1/3$ and $2\chi=1$. We then perform simulations with $\lambda_1\neq0$. In Fig.~\ref{fig:W2lambda2} we plot $W^2(t,L)$ as a function of time and various system sizes both in log-log and in a redressed plot (inset); the latter is a stringent test of the scaling behavior. Our measurement gives $2\chi/z\simeq 0.28$. We then plot the saturated value of the interfacial width in Fig.~\ref{fig:scalingchi} both for $\lambda_1=0$ and $\lambda_1=2$, finding that $2\chi=1$ for $\lambda_1=0$ and $2\chi\simeq 0.78$ for $\lambda_1=2$. The values of the measured scaling exponents should be compared with those obtained from the one-loop RG analysis, Eq.~(\ref{eq:critical-exp}), which gives $2\chi/z\simeq 1/4$ and $2\chi =2/3$ in $d=1$. The
agreement is very good, given that the RG predictions are obtained to first order in $\varepsilon$, whereas $\varepsilon = 1$ in our simulations. These numerical results offer strong evidence for the presence of a perturbatively accessible fixed point for the $\abs{\bq}$KPZ equation.

Our final task is to evidence that the $\abs{\bq}$KPZ nonlinearity is indeed generically present for the dynamics of interfaces in phase-separated active systems. To do this, we consider continuum scalar field theories for active phase separation ~\cite{Wittkowski14,tjhung2018cluster,thomsen2021periodic}. These are generalizations of Model B~\cite{hohenberg1977theory,chaikin2000principles,bray2001interface}, the standard large-scale description of phase separation in diffusive passive fluids (without momentum conservation). Like Model B they address the dynamics of a conserved scalar density field $\phi$ in $d+1$ dimensions which switches steeply from positive to negative values on crossing the interface from liquid to vapor. Unlike Model B, the active versions take into account that detailed balance is broken at the microscale. The ensuing minimal theory is called AMB+~\cite{nardini2017entropy,tjhung2018cluster}, and includes all terms that break detailed balance up to order $\mathcal{O}(\nabla^4,\phi^2)$~\cite{tjhung2018cluster}: 
\qq
\p_t\phi&=&-\nabla\cdot\left(\mathbf{J}+\sqrt{2DM}\bm{\Lambda}\right),\label{eq:AMB+}\\
 {\bf J}/M &=&-\nabla \mu_\lambda  + \zeta (\nabla^2\phi)\nabla\phi,\label{eq:AMB+J}\\
\mu_\lambda[\phi] &=& \frac{\delta \mathscr{F}}{\delta\phi} +\lambda|\nabla\phi|^2.\label{eq:AMB+mu}
\qqq
Here $\mathscr{F} = \int d\bfr \,\left[f(\phi) +\frac{K(\phi)}{2}|\nabla\phi|^2\right]$, $f(\phi)$ is a double-well local free energy,
and $\bm{\Lambda}$ is a vector of zero-mean, unit-variance, Gaussian white noises. Standard Model B is recovered at vanishing activity
($\lambda=\zeta=0$), unit mobility $M=1$ and constant noise level $D$~\cite{hohenberg1977theory}. Here we shall retain the choice $M=1$, constant $D$ and further assume for simplicity that $K$ is also constant. Our analysis below assumes that the interfacial tensions determining the Ostwald process~\cite{tjhung2018cluster} and the relaxation of capillary waves~\cite{fausti2021}, which in active systems can differ, are both positive so that the system undergoes bulk phase separation rather than microphase separation. 

The linear description of capillary waves, Eq.~(\ref{eq:linearLangevin}), has been classically derived for passive fluids starting from Model B~\cite{bray2001interface} and, more recently, for active fluids starting from AMB+~\cite{fausti2021}. This is done by assuming that $\phi$ evolves quasistatically with respect to fluctuations of the interfacial height, equivalent to an {\em ansatz} $\phi(\bx,y,t)=\varphi(y-\hat{h}(\bx,t))$, which is exact at leading order in $h$ and $\bq$~\cite{fausti2021}. As detailed in~\cite{supp}, we extend this procedure to obtain the nonlinear terms that correct Eq.~(\ref{eq:linearLangevin}):
\begin{widetext}
\begin{align}\label{eq:effective-h-equation}
\sum_{n=0}^{\infty} \int_{\bq_1,\bx,\bx_1} 
(-1)^{n}
 \frac{(\hat{h}(\bx)-\hat{h}(\bx_1))^n}{2n!}
\abs{\bq_1}^n \left\{
A_n(\bq_1) \partial_t \hat{h}(\bx)
+ \zeta   D_n(\bq_1) \bG^2 \hat{h}
 \right\}e^{-i \bq \cdot \bx_1 - i \bq_1 \cdot \bx_1 + i \bq_1 \cdot \bx } 
   =
   -\sigma_\lambda \bq^2 h_\bq +\chi_\bq,
\end{align}
\end{widetext}
where the Stratonovich convention is used, expressions for the $h$-independent factors $A_n(\bq), D_n(\bq), \sigma_\lambda$ are given in~\cite{supp}, and $\chi_\bq$ can be found as a sum of Gaussian noises~\cite{supp}. $A_n(\bq)$ and $D_n(\bq)$ have been defined to make clear the reading of the dimension of the nonlinearities, given the fact that the dimension of $A_n(\bq) \partial_t \hat{h}(\bx)$ and $D_n(\bq) \bG^2 \hat{h}$ is at least $\mathcal{O}(\abs{\bq}^2 \hat{h})$. To leading order in $h$ and $\bq$, Eq.~(\ref{eq:effective-h-equation}) reduces to Eq.~(\ref{eq:linearLangevin}), where the interfacial tension $\sigma$ is proportional to the capillary waves interfacial tension $\sigma_{\rm cw}$~\cite{fausti2021}.

Notably, the $\lambda_1$-term of Eq.~(\ref{eq:qKPZ}) is not present at bare level in Eq.~(\ref{eq:effective-h-equation}). Furthermore, Eq.~(\ref{eq:effective-h-equation}) contains nonlinearities in the form of Eq.~(\ref{eq:Ansatz}) with $g$ singular. In order to perform the RG analysis of Eq.~(\ref{eq:effective-h-equation}), we transform it to the Ito convention in~\cite{supp}, following a standard procedure~\cite{Gardiner:1985,lau2007state,cates2022stochastic}. We then find the canonical dimension of the nonlinear terms and show that the singular nonlinearities in Eq.~(\ref{eq:effective-h-equation}) are all RG irrelevant~\cite{supp}. Moreover, by computing one-loop diagrams we further show in~\cite{supp} that $\lambda_1\neq0$ is generated under RG from the nonlinearities of Eq.~(\ref{eq:effective-h-equation}). These results strongly suggest that the liquid-vapor interface of AMB+ belongs to the $\abs{\bq}$KPZ universality class. However a complete proof would require us to derive the analog of Eq.~(\ref{eq:effective-h-equation}) allowing for a dependence of $\phi$ on interfacial curvature, and then show that no relevant singular nonlinearity is generated by the associated RG flow. This goes beyond the scope of this Letter. Preliminary numerical simulations of AMB+, which will be presented elsewhere, do however indicate that the interface roughens accordingly to $W^2(t)\sim t^{0.28}$ as expected from the $\abs{\bq}$KPZ universality class.

% ==============================================================================
In conclusion, we have introduced a minimal field theory, termed the $\abs{\bq}$KPZ equation, to describe the roughening of interfaces in nonequilibrium phase-separated systems lacking momentum conservation. The $\abs{\bq}$KPZ equation differs from the standard description of roughening interfaces (either EW, KPZ or their conserved counterparts) because diffusive fluxes in the bulk cause the interfacial dynamics to be nonlocal in real space. We discovered a nontrivial fixed point of the RG flow for interfacial dimension $d< 2$. This should control active, phase-separated interfaces in $d+1 = 2$ bulk dimensions. We characterized this new universality class, computing its scaling exponents by one-loop RG and numerical simulations.
We finally gave evidence that the $\abs{\bq}$KPZ class includes interfacial roughening in phase-separated active systems, by explicitly deriving the effective interface equation from a scalar active field theory for the particle density in $d+1$ bulk dimensions, and showing that the $\abs{\bq}$KPZ nonlinearity thereby emerges. 

Previous studies of roughening of the liquid-vapor interface in active particle models concluded that $z\sim 2$ as in the Edwards-Wilkinson universality class~\cite{wysocki2016propagating,lee2017interface,patch2018curvature}. We speculate that such disagreement might arise if these particle models undergo bubbly phase separation~\cite{tjhung2018cluster} instead of bulk phase separation as addressed here. (Notably, in~\cite{patch2018curvature}, vapor bubbles are indeed  visible in the liquid phase). Local events in which a vapor bubble ``pops'' through the interface do not conserve the total height $\int_{\bx} ~\hat{h}$, and also create overhangs, evading our description -- at least on scales smaller than the largest bubbles present.  More work is needed to clarify this aspect. Beyond active systems, our results might also describe the roughening of interfaces in other nonequilibrium phase-separating systems, such as granular materials~\cite{oyarte2013phase}. 

\begin{acknowledgments}
CN thanks F. Caballero for a discussion in the initial stages of this work. GF was supported by the CEA NUMERICS program, which has received funding from the European Union's Horizon 2020 research and innovation program under the Marie Sklodowska-Curie grant agreement No 800945. BD acknowledges the support from the French ANR through the project NeqFluids (grant ANR-18-CE92-0019). Work funded in part by the European Research Council under the Horizon 2020 Programme, ERC grant agreement number 740269 and by the National Science Foundation under Grant No. NSF PHY-1748958, NIH Grant No. R25GM067110 and the Gordon and Betty Moore Foundation Grant No. 2919.02. MEC is funded by the Royal Society. CN acknowledges the support of an Aide Investissements d'Avenir du LabEx PALM (ANR-10-LABX-0039-PALM). 
\end{acknowledgments}

% ==============================================================================

\bibliographystyle{apsrev4-1}
\bibliography{biblio.bib}

\clearpage

\end{document}